\documentclass{PoS}

\title{Relation between scattering amplitude and
Bethe-Salpeter wave function in quantum field theory}

\ShortTitle{Relation between scattering amplitude and
Bethe-Salpeter wave function in quantum field theory}

\author{\speaker{Takeshi Yamazaki}$^{a,b}$\thanks{E-mail: yamazaki@het.ph.tsukuba.ac.jp}\hspace{2mm}  and Yoshinobu Kuramashi$^{b}$\vspace{2mm}
\\
\llap{$^a$}Faculty of Pure and Applied Sciences,
University of Tsukuba, Tsukuba, Ibaraki 305-8571, Japan\\
   \llap{$^b$}
Center for Computational Sciences, University of Tsukuba,
Tsukuba, Ibaraki 305-8577, Japan}

\abstract{
We discuss an exact relation between the two-particle scattering
amplitude and the Bethe-Salpeter (BS) wave function inside 
the interaction range in quantum field theory.
In the relation the reduced BS wave function defined by the BS wave function
plays an essential role. Through the relation the on-shell and half off-shell
amplitudes can be calculated.
We also show that the solution of Schr\"odinger equation 
with the effective potential determined from the BS wave function
gives a correct on-shell scattering amplitude 
only at the momentum where the effective potential is determined.
Furthermore we discuss a derivative expansion of the reduced BS wave function 
and a condition to obtain results independent of the interpolating operators
in the time-dependent HALQCD method.
}

\FullConference{The 36th Annual International Symposium on Lattice Field Theory - LATTICE2018\\
		22-28 July, 2018\\
		Michigan State University, East Lansing, Michigan, USA.}

\begin{document}

\section{Introduction}

The finite volume method~\cite{Luscher:1986pf,Luscher:1990ux} 
is utilized in various lattice studies for hadron scatterings.
In the method the scattering phase shift $\delta(k)$
is evaluated from the finite volume formula.
The formula relates $\delta(k)$ in the infinite volume to
the relative momentum of two particles $k^2$ on finite volume.
The derivation of the formula was based on a relation between
the two-particle wave function 
outside the interaction range $R$ and $\delta(k)$
in quantum mechanics~\cite{Luscher:1990ux}.
The same formula was obtained from a similar discussion
with the Bethe-Salpeter (BS) wave function in 
quantum field theory~\cite{Lin:2001ek,Aoki:2005uf}.
The relation between the BS wave function outside $R$
and the scattering amplitude is well understood in the finite volume method.

On the other hand,
the relation between the BS wave function inside $R$ and 
the scattering amplitude is not well known.
Only the HALQCD method~\cite{Aoki:2009ji} was proposed
based on a discussion in quantum mechanics,
which is a method to evaluate $\delta(k)$ with
an effective potential determined from the BS wave function inside $R$.
In lattice QCD calculation of the two-nucleon channels,
HALQCD method and the direct calculation of bound state energy
give qualitatively different results.
The reason of the inconsistency has not been understood at present,
though several possible reasons are suggested.

In this report, we discuss an exact relation between 
the BS wave function inside $R$ in the infinite volume
and the on-shell scattering amplitude in quantum field theory.
Based on this relation, we show that the
correct scattering phase shift is obtained from the effective potential
only at the momentum where the effective potential is determined.
At other momenta, however, the scattering amplitude obtained from
the effective potential disagrees with that from the exact relation.
Furthermore, we discuss a derivative expansion of the potential 
and a condition to obtain results independent of the interpolating operators 
in the time-dependent HALQCD method~\cite{HALQCD:2012aa}.
These discussions could be useful to understand property of hadron scatterings
and also the inconsistency between the two methods in 
the two-nucleon lattice QCD calculations.
All the results in this report have been already published in
the two papers~\cite{Yamazaki:2017gjl,Yamazaki:2018qut}.

\section{BS wave function inside interaction range}

In this report, we follow the definitions in 
Refs.~\cite{Lin:2001ek,Aoki:2005uf}.
We consider an S-wave 
scattering of distinguishable spinless particles.
The relative momentum of the two particles $k$ is determined from
the two-particle energy $2E_k = 2\sqrt{m^2+k^2}$ where $m$ is the mass
of the particles.
The energy is below the inelastic threshold $2E_k < 4m$.
The BS wave function of the two-particle
in the infinite volume is defined by
\begin{equation}
\phi(\vec{x};\vec{k}) = \langle 0 | \pi_1(\vec{x}/2)\pi_2(-\vec{x}/2) |
\hat{\pi}_1(\vec{k})\hat{\pi}_2(-\vec{k}); {\rm in}\rangle,
\label{eq:defphixk}
\end{equation}
where $\pi_i$ is an interpolating operator of the $i$-th scalar particle 
($i=1,2$),
and $|\hat{\pi}_1(\vec{k})\hat{\pi}_2(-\vec{k}); {\rm in}\rangle$ is
an asymptotic two-particle state with the relative momenta $\vec{k}$
and $-\vec{k}$.
We omit $t$ dependence of $\phi(\vec{x};\vec{k})$, because it can be 
expressed by an overall factor $e^{i 2E_k t}$.

Through the LSZ reduction formula,
$\phi(\vec{x};\vec{k})$ is written by the half off-shell scattering amplitude
$H(p;k)$~\cite{Lin:2001ek,Aoki:2005uf} with $k = |\vec{k}|$ as
\begin{equation}
\phi(\vec{x};\vec{k}) = e^{i\vec{k}\cdot\vec{x}} +
\int\frac{d^3p}{(2\pi)^3}\frac{H(p;k)}{p^2-k^2-i\epsilon}
e^{i\vec{p}\cdot\vec{x}} .
\label{eq:def_phixk}
\end{equation}
In this expression, unnecessary overall factors are ignored.
$H(p;k)$ is defined in the LSZ reduction formula, which is given by
the Fourier transformation of the four-point Green function as,
\begin{eqnarray}
&&
e^{-i{\bf q}\cdot{\bf x}}
\frac{-i\sqrt{Z} }{-{\bf q}^2 + m^2 - i\varepsilon}\frac{8E_p E_k}{E_p + E_k}
H(p;k) =
\nonumber\\
&& \int d^4zd^4y_1d^4y_2 
K({\bf p},{\bf z}) K(-{\bf k_1},{\bf y_1}) K(-{\bf k_2},{\bf y_2})
\langle 0 | T[ \pi_1({\bf z})\pi_2({\bf x})\pi_1({\bf y_1})\pi_2({\bf y_2})
| 0 \rangle,
\label{eq:LSZ}
\end{eqnarray}
where 
$K({\bf p},{\bf z}) = ie^{i{\bf p}\cdot{\bf z}}
(-{\bf p}^2 + m^2)/\sqrt{Z}$,
with $Z$ the renormalization factor of the operator $\pi_i$.
The bold faced momenta and coordinates are four-dimensional vectors.
Three of the four momenta, ${\bf p}, {\bf k_1},$ and ${\bf k_2}$, are 
on-shell,
while ${\bf q} = (2E_k - E_p, -\vec{p})$ is generally off-shell.
The on-shell scattering amplitude $H(k;k)$ 
is written by $\delta(k)$ as
\begin{equation}
H(k;k) = \frac{4\pi}{k} e^{i\delta(k)}\sin\delta(k) .
\label{eq:hkk}
\end{equation}
In the S-wave BS wave function $\phi(x;k)$, 
the first term in Eq.~(\ref{eq:def_phixk})
is replaced by its S-wave component $j_0(kx)$ which is 
the spherical Bessel function of $l = 0$.

We define the reduced BS wave function
given by $\phi(x;k)$ as,
\begin{equation}
h(x;k) = (\Delta + k^2) \phi(x;k) .
\label{eq:def_hxk}
\end{equation}
An important assumption of $h(x;k)$ is that
$h(x;k) = 0$ in the outside region of the interaction range $R$
except for the exponential tail.
This property is similar to a potential in quantum mechanics.
Using Eq.~(\ref{eq:def_phixk}) we can see that $h(x;k)$ is
directly related to $H(p;k)$ as
\begin{equation}
h(x;k) =
-\int\frac{d^3p}{(2\pi)^3} H(p;k)e^{i\vec{p}\cdot\vec{x}} .
\label{eq:hxk_Hpk}
\end{equation}

The Fourier transformation of $h(x;k)$ gives 
the half off-shell amplitude,
\begin{equation}
H(p;k) = -\int d^3x\, h(x;k) e^{-i\vec{p}\cdot\vec{x}} .
\label{eq:def_hpk}
\end{equation}
At the on-shell $p=k$,
the scattering phase shift $\delta(k)$ is obtained
from the on-shell scattering amplitude $H(k;k)$ as,
\begin{equation}
H(k;k) = -\int d^3x\, h(x;k) e^{-i\vec{k}\cdot\vec{x}} =
\frac{4\pi}{k}e^{i\delta(k)}\sin\delta(k) ,
\label{eq:def_hkk}
\end{equation}
where Eq.~(\ref{eq:hkk}) is used in the last equality.
This is an exact relation between the BS wave function inside $R$
and the on-shell scattering amplitude in quantum field theory,
because $h(x;k)$ given by $\phi(x;k)$ has non-zero value only in the inside
region of the interaction range.
We will call the relation the fundamental relation in this report.
The fundamental relation insists that
$h(x;k)$ plays an essential role to calculate $\delta(k)$ 
in quantum field theory, when we use the BS wave function inside $R$.
Although this relation
is not explicitly written, it was used to
show the relation between $\delta(k)$ and $\phi(x;k)$ in
$x > R$ in Ref.~\cite{Aoki:2005uf}.
One can extend the formula to the one on finite volume.
The first calculation for the on-shell and half off-shell scattering amplitudes 
with the extended formula was reported in Ref.~\cite{Namekawa:2017sxs}.

\section{Fundamental relation in quantum mechanics}
\label{sec:schrodinger_eq}

In this section we will compare the on-shell scattering amplitude
obtained from the fundamental relation and the one from
the solution of Schr\"odinger equation with an effective potential
given by the BS wave function.

We define the effective potential $V(x;k)$ by $h(x;k)$ 
and $\phi(x;k)$~\cite{Aoki:2005uf} as,
\begin{equation}
V(x;k) = 
\frac{1}{m}\frac{h(x;k)}{\phi(x;k)}\ \  (x \le R) .
\label{eq:def_vxk}
\end{equation}
We assume $V(x;k) = 0$ in $x > R$.
$V(x;k)$ may diverge, if $\phi(x;k)$ has a node in $x \le R$.

In the leading order HALQCD method~\cite{Aoki:2009ji},
$V(x;k)$ is regarded as a potential in quantum mechanics.
In the method the scattering phase shift $\overline{\delta}(p)$ is 
determined from the solution of Schr\"odinger equation 
with a given momentum $p$ (in general $p \ne k$) below the threshold.
Schr\"odinger equation is given by
\begin{equation}
(\Delta + p^2) \overline{\phi}(x;p) = 2\mu V(x;k) \overline{\phi}(x;p),
\label{eq:schrodinger}
\end{equation}
where $\overline{\phi}(x;p)$ is the solution
and $\mu$ is the reduced mass $\mu = m/2$.

As explained in textbook of quantum mechanics 
(see for example Ref.~\cite{Sakurai:1993zz}),
the scattering amplitude $f(p)$ is obtained from Schr\"odinger equation
Eq.~(\ref{eq:schrodinger}) as
\begin{equation}
f(p) = -\frac{2\mu}{4\pi}\int d^3x\, V(x;k) \overline{\phi}(x;p) 
e^{-i\vec{p}\cdot\vec{x}}=
-\frac{1}{4\pi} \int d^3x \frac{h(x;k)}{\phi(x;k)} \overline{\phi}(x;p) 
e^{-i\vec{p}\cdot\vec{x}} ,
\label{eq:fp_phi}
\end{equation}
where the definition of $V(x;k)$ Eq.~(\ref{eq:def_vxk})
is used in the last equality.
In the equation, it is assumed that
$f(p)$ can be also written by $\overline{\delta}(p)$ as
\begin{equation}
f(p) = \frac{e^{i\overline{\delta}(p)}\sin\overline{\delta}(p)}{p} .
\label{eq:fp_delta}
\end{equation}
In the following, we compare the scattering phase shifts
obtained from Schr\"odinger equation and the fundamental relation,
$\overline{\delta}(p)$ and $\delta(p)$, in two cases:
$p = k$ and $p \ne k$,
since the effective potential $V(x;k)$ is 
defined at $k$ as in Eq.~(\ref{eq:def_vxk}).

At $p=k$,
the above Schr\"odinger equation reduces
to the definition of $h(x;k)$ in Eq.~(\ref{eq:def_hxk}).
In this case
$\overline{\phi}(x;k) = \phi(x;k)$ in Eq.~(\ref{eq:fp_phi}).
Thus, $f(k)$ in Eq.~(\ref{eq:fp_phi}) is written by 
$\delta(k)$ using the fundamental relation Eq.~(\ref{eq:def_hkk}) as,
\begin{equation}
f(k) = -\frac{1}{4\pi} \int d^3x\, h(x;k)e^{-i\vec{k}\cdot\vec{x}} = 
\frac{1}{4\pi}H(k;k) =
\frac{e^{i\delta(k)}\sin\delta(k)}{k} .
\end{equation}
Comparing the result with $f(k)$ in Eq.~(\ref{eq:fp_delta}), 
it is confirmed $\overline{\delta}(k) = \delta(k)$.
Therefore, the same scattering phase shift is obtained from
Schr\"odinger equation and the fundamental relation
at the momentum $k$ where $V(x;k)$ is defined.

On the other hand, in the $p \ne k$ case, the two scattering phase
shifts do not agree.
It is because generally $\overline{\phi}(x;p) \ne \phi(x;k)$
in this case.

\section{Expansion of reduced BS wave function}
\label{sec:expansion_h}

The reduced BS wave function $h(x;k)$ is expanded by derivatives
in HALQCD method~\cite{Aoki:2009ji} as,
\begin{equation}
h(x;k) = \sum_n^\infty V_n(x) \Delta^n \phi(x;k).
\label{eq:derivative_exp}
\end{equation}
The expansion coefficient $V_n(x)$ is independent of $k$,
if the number of the term in the expansion is infinite~\cite{Kawai:2017goq}.

In practical calculation, the expansion is truncated to some order.
In this case the convergence of the expansion is unclear,
because it is not a systematic expansion.
Furthermore $V_n(x)$ depends on $k$ in contrast to the expansion
with the infinite terms.
In the next section we will discuss uncertainties caused by the truncation 
in the time-dependent HALQCD method~\cite{HALQCD:2012aa},
because the $k$ independence of $V_n(x)$ is a theoretical base of
the method.

It is easy to see the $k$ dependence of $V_n(x)$.
For example, when the expansion is truncated to the two terms,
$h(x;k)$ is given by
\begin{eqnarray}
h(x;k)
&=&V_0(x)\phi(x;k) + V_1(x)\Delta \phi(x;k)\\
&=&V_0(x)\phi(x;k) + V_1(x)(h(x;k)-k^2\phi(x;k)) .
\end{eqnarray}
The definition of $h(x;k)$ in Eq.~(\ref{eq:def_hxk}) is used 
in the last equality.
By solving simultaneous equations with
given $h(x;k)$ and $\phi(x;k)$ at two momenta,
one can see that $V_0(x)$ and $V_1(x)$ depend on the two momenta.
The obtained coefficients give the correct $h(x;k)$
at the two input momenta, while they do not in other momenta.
This is similar to the situation of the effective potential as
discussed in the previous section.
The effective potential is determined from $h(x;k)$ and $\phi(x;k)$
at one momentum $k$, so that the correct scattering amplitude
is obtained only at $k$.

\section{Truncated expansion in time-dependent HALQCD method}
\label{app:thal}

The time-dependent HALQCD method~\cite{HALQCD:2012aa} was proposed 
based on the derivative expansion
to obtain $k$ independent $V_n(x)$ by solving simultaneous equations
of two-particle correlation functions on the lattice.
As discussed in the previous section, such $V_n(x)$ cannot be obtained
in practical calculation.
In this section 
we discuss a condition to obtain $V_n(x)$ that do not depend on
the choice of the interpolating operators.

The correlation function on the lattice $C_i(x,t)$
is expanded by the two-particle states with the discrete momentum $k_\alpha$ as,
\begin{eqnarray}
C_i(x,t) &=& \langle 0| \pi(x,t)\pi(0,t)\Omega_i|0\rangle
= \sum_{\alpha=0}^{N_\alpha} A_{i\alpha}(t) \phi_\alpha(x),
\end{eqnarray}
where $\phi_\alpha(x)$ corresponds to $\phi(x;k_\alpha)$
and $A_{i\alpha}(t) = B_{i\alpha} e^{-E_\alpha t}$ with 
$B_{i\alpha} = \langle \pi\pi;k_\alpha|\Omega_i|0\rangle$ and 
$E_\alpha^2 = 4 (m^2+k_\alpha^2)$.
$\Omega_i$ $(i = 0, \cdots, N_\Omega)$ 
is the $i$-th two-particle operator at $t = 0$.
We assume that $N_\alpha+1$ states contribute to $C_i(x,t)$
in the $t$ region.
The number of the states decreases as $t$ increases, because contributions of 
higher energy states are exponentially suppressed by $t$.
The sum of the reduced BS wave function $h_\alpha(x) = h(x;k_\alpha)$
is obtained from $C_i(x,t)$ as
\begin{equation}
(\Delta + f(t,m))C_i(x,t) =
\sum_{\alpha=0}^{N_\alpha} A_{i\alpha}(t) h_\alpha(x) ,
\label{eq:thal_base}
\end{equation}
where $f(t,m)$ satisfies
$f(t,m) A_{i\alpha}(t) = k^2_\alpha A_{i\alpha}(t)$.

The time-dependent HALQCD method employs the truncated expansion
using $N$ derivatives with $k$ independent $V_n(x)$,
although the truncation causes the $k$ dependence of $V_n(x)$
as explained in the previous section.
Thus, the right-hand side of Eq.~(\ref{eq:thal_base}) is given by
\begin{equation}
\sum_{\alpha=0}^{N_\alpha} A_{i\alpha}(t) h_\alpha(x) =
\sum_{\alpha=0}^{N_\alpha} A_{i\alpha}(t) \sum_{n=0}^N V_{n}(x) \Delta^n \phi_\alpha(x) =
\sum_{n=0}^N V_n(x) \Delta^n C_i(x,t) ,
\label{eq:simuleq_org}
\end{equation}
where the summations for $\alpha$ and $n$ are exchanged in the
last equality because of the $k$ independence of $V_n(x)$.
For convenience, we define matrices $A(t)$ and $M(x,t)$ and vectors
$h(x)$ and $V(x)$, whose components are $A_{i\alpha}(t)$, 
$M_{in}(x,t) = \Delta^n C_i(x,t)$, $h_\alpha(x)$, and $V_n(x)$,
respectively.
Using the matrices and vectors, Eq.~(\ref{eq:simuleq}) is expressed by
\begin{equation}
M(x,t) V(x) = A(t) h(x) .
\label{eq:simuleq}
\end{equation}

In order to solve the simultaneous equations Eq.~(\ref{eq:simuleq}),
the inverse of $M(x,t)$ is calculated in $N = N_\Omega$.
Thus, $V(x)$ is given by
\begin{equation}
V(x) = (M(x,t))^{-1} A(t) h(x) .
\end{equation}
Since $A(t)$ depends on the operators,
$A(t)$ should vanish to obtain operator independent $V_n(x)$.
However, if $N_\alpha \ne N$, $A(t)$ does not vanish,
because $A(t)$ does not have an inverse. 
$(M(x,t))^{-1}$ cannot be decomposed into
two inverse matrices for $A(t)$ and $\Phi(x)$, where 
$\Phi_{\alpha n}(x)=\Delta^n\phi_\alpha(x)$ and
$M(x,t) = A(t)\Phi(x)$.
In this case $V(x)$ is a function of 
$A_{n\alpha}(t), h_\alpha(x)$, and $\Delta^n \phi_\alpha(x)$,
so that $V_n(x)$ depends on the choice of the interpolating operator
in $C_i(x,t)$.

On the other hand, in the case of $N_\alpha = N$,
$A(t)$ disappears as,
\begin{equation}
V(x) = (M(x,t))^{-1} A(t) h(x)
= (\Phi(x))^{-1} h(x).
\end{equation}
Therefore, $N_\alpha = N$ is a condition to obtain
the operator independent $V_n(x)$ from the time-dependent HALQCD method.

In order to satisfy $N_\alpha = N$, in general
$C_i(x,t)$ in a large $t$ region is necessary,
where higher energy states than the $N_\alpha+1$ states 
must be sufficiently suppressed.
It might be also possible to 
adopt operators which strongly couple to the states of 
$\alpha = 0, \cdots, N_\alpha$.
These conditions are the same as
in the generalized eigenvalue problem~\cite{Luscher:1990ck}
to obtain the two-particle energies.
In the time-dependence HALQCD method the condition may be more severe,
because it must be satisfied in all $x$.

Even if the operator independent $V_n(x)$ is obtained,
it gives the correct $h_\alpha(x)$ only in the momenta $k_\alpha$,
because $V_n(x)$ should have the momentum dependence due to
the truncation of the derivative expansion.
However, the momenta cannot be determined from the time-dependent HALQCD method,
so that other method is required to specify the momenta
where the correct scattering amplitudes are obtained.

\section{Summary}

We have presented an exact relation between the BS
wave function inside the interaction range and the 
scattering amplitude in quantum field theory.
This relation gives not only the on-shell amplitude but also
the half off-shell amplitude.
The reduced BS wave function plays an essential role in
the relation, which is defined by the BS wave function.

Using the relation, we have presented that
Schr\"odinger equation with the effective potential 
determined from the reduced BS wave function
gives the same scattering phase shift as in quantum field theory
only at the momentum where the effective potential is defined.
In other momenta the two scattering phase shifts differ in general.

We have also discussed that the truncated expansion of the reduced BS wave 
function causes the momentum dependence of the expansion coefficients.
Furthermore it is discussed uncertainties caused by the truncation 
in the time-dependent HALQCD method and
a condition to obtain results independent of the interpolating operators.

These discussions could be helpful to understand the current situation
of the two-nucleon calculations in lattice QCD.

\section*{Acknowledgements}
This work is supported in part by Japan Society for the Promotion of Science (JSPS) Grants-in-Aid for Scientific Research for Young Scientists (A) No. 16H06002.

\bibliography{reference}

\end{document}